\documentclass[conference]{IEEEtran}
\usepackage{amssymb,amsfonts,amsmath,amsthm,amscd,dsfont,mathrsfs}
\DeclareMathAlphabet{\mathpzc}{OT1}{pzc}{m}{it}

\newtheorem{propo}{Proposition}[section]
\newtheorem{lemma}[propo]{Lemma}

\newcommand{\<}{\langle}
\renewcommand{\>}{\rangle}

\newcommand{\reals}{{\mathds R}}

\newcommand{\naturals}{{\mathds N}}

\def\E{\mathds E}
\def\1{\mathds 1}
\def\de{{\rm d}}

\def\me{{\nu}}
\def\mh{\hat{\nu}}
\def\normeq{{\cong}}
\def\ph{\hat{\phi}}
\def\pe{\phi}
\def\htau{\hat{\tau}}
\def\Var{{\rm Var}}
\def\F{{\sf F}}
\def\G{{\sf G}}
\def\tY{\widetilde{Y}}
\def\argmin{{\rm argmin}}

\begin{document}

\title{Message Passing Algorithms for Compressed Sensing:
I. Motivation and Construction}

\author{\IEEEauthorblockN{David L. Donoho}
\IEEEauthorblockA{Department of Statistics\\
Stanford University
}
\and
\IEEEauthorblockN{Arian Maleki}
\IEEEauthorblockA{Department of Electrical Engineering\\
Stanford University}
\and
\IEEEauthorblockN{Andrea Montanari}
\IEEEauthorblockA{Department of Electrical Engineering\\
and Department of Statistics\\
Stanford University
}}

\maketitle

\begin{abstract}
In a recent paper, the authors proposed a new class of low-complexity
iterative thresholding algorithms for reconstructing sparse signals
from a small set of linear measurements \cite{DMM}.
The new algorithms are broadly referred to as 
AMP, for \emph{approximate message passing}.
This is the first of
two conference papers describing the derivation of
these algorithms, connection with the related literature,
extensions of the original framework, and new empirical evidence.

In particular, the present paper outlines the derivation
of AMP from standard sum-product belief propagation,
and its extension in several directions. We also discuss
relations with formal calculations based on statistical mechanics
methods.
\end{abstract}

\IEEEpeerreviewmaketitle

\section{Introduction}

Let $s_o$ be a vector in $\reals^N$.
We observe $n<N$ linear measurements of this vector through the matrix $A$,
$y= As_o$. The goal is to recover $s_o$ from $(y,A)$.
Although the system of equations is underdetermined, 
the underlying signal can still be recovered exactly or approximately
if it is `simple' or `structured' in an appropriate sense. 
A specific notion of `simplicity' postulates that
$s$ is  exactly or approximately sparse. 

The $\ell_1$ minimization, also known as the basis pursuit 
\cite{ChDoSa98}, has attracted attention for its success in solving such
underdetermined systems. It consists in solving 
the following optimization problem:
\begin{eqnarray} \label{basispursuit}
\mbox{minimize} \; \|s\|_1  \, ,\;\;\;\;\;\; \mbox{subject to}\;\;\; As = y\, .
\end{eqnarray}
The solution of this problem can be obtained through generic
linear programming (LP) algorithms. While LP has polynomial
complexity, standard LP solvers are too complex for use in large scale
applications, such as magnetic resonance imaging and seismic data analysis. 
Low computational complexity of iterative thresholding algorithms 
has made them an appealing choice for such applications. 
Many variations of these approaches have been proposed. The interested 
reader is referred to \cite{MaDo09} for a survey and detailed comparison. 
The final conclusion of that paper is 
rather disappointing: optimally tuned 
iterative thresholding algorithms have a significantly worse 
sparsity-undersampling tradeoff than basis pursuit. 

Recently \cite{DMM}, we proposed an algorithm that appears to 
offer the best of both worlds: 
the low complexity of iterative thresholding algorithm, 
and the reconstruction power of the basis pursuit \cite{DMM}. 
This algorithm is in fact an instance of a broader family of algorithms, 
that was called AMP, for approximate message passing, in \cite{DMM}. 
The goal of this paper is  to justify AMP by
applying  sum-product belief propagation for a suitable 
joint distribution  over the variables $s_1, s_2, \ldots, s_N$. 

The paper is organized as follows: In Section \ref{Sec:Notation} 
we explain the notations used in this paper. We then derive the 
AMP algorithm associated the basis pursuit problem in Section 
\ref{sec:AMPhard}. In Section  \ref{sec:AMPsoft}, we consider the AMP 
for the basis pursuit denoising (BPDN) or Lasso problem. We will also  
generalize the algorithm to the Bayesian setting where the 
distribution of the elements of $s_o$ is known, 
in Section \ref{sec:AMPBayes}. Finally we will explain the
connection with formal calculations based on 
non-rigorous statistical mechanics methods in Section \ref{sec:Related}.

Due to space limitations, proofs are omitted and can be found in a longer
version of this paper \cite{DMMFuture}.
%
%
\section{Notations}\label{Sec:Notation}

 The letters $a,b,c,\dots$ denote indices in $[n] \equiv\{1,\dots,n\}$ and
$i,j,k,\dots$ represent indices in $[N]\equiv\{1,\dots,N\}$.
The $a,i$ element of the matrix $A$ will be indicated as $A_{ai}$.
The elements of the vectors $y$, $s$, $x$, and $s_o$ are indicated by
$y_a$, $s_i$, $x_i$, and $s_{o,i}$ respectively.

The ratio $\delta= n/N$ is a measure of 
indeterminacy of the system of equations.
Whenever we refer to the large system limit we consider 
the case where $N , n\to \infty$ with $\delta$ fixed.
In this limit the typical entry of $A$ should scale as $1/\sqrt{n}$.
In the concrete derivation, for the sake of simplicity we assume
that $A_{ai}\in \{+ 1/\sqrt{n},- 1/\sqrt{n} \}$. This assumption is
not crucial, and only simplifies the calculations. Although the 
algorithms are developed from the large system limit,
 in practice, they perform well even
in the medium size problems with `just' 
thousands of variables and hundreds of measurements \cite{DMM2}.
%
%
\section{AMP for the Basis Pursuit}
\label{sec:AMPhard}

In this section we consider the the basis pursuit problem as defined in Eq. (\ref{basispursuit}). The derivation of AMP proceeds in
4 steps:

\noindent\emph{(1)} Construct a joint distribution
over $(s_1,\dots,s_N)$,
parameterized by $\beta\in\reals_+$, associated with the problem
of interest and write down the corresponding sum-product algorithm.

\noindent\emph{(2)} Show, by a central limit theorem argument, that
for the large system limit, the sum-product messages can well be approximated by the
families with two scalar parameters. Derive the update rules for these
parameters.

\noindent\emph{(3)} 
Take the limit $\beta\to\infty$ and get the appropriate rules
for basis pursuit problem.

\noindent\emph{(4)}  Approximate the message passing rules for the large system limit. The resulting algorithm is AMP.
%
%
%
\subsection{Construction of the graphical model}
We consider the following joint probability distribution over the
variables $s_1, s_2, \ldots s_N$
\begin{align}
\mu(\de s)= \frac{1}{Z} \prod_{i=1}^N \exp{(- \beta |s_i|)} \prod_{a=1}^{n} \delta_{\{y_a= (As)_a\}}\, .\label{eq:JointDistribution}
\end{align}
Here $\delta_{\{y_a= (As)_a\}}$ denotes a Dirac distribution
on the hyperplane $y_a = (Ax)_a$.  Products of such distributions
associated with distinct hyperplanes yield a well defined measure.
As we let $\beta\to\infty$, the mass of   $\mu$
concentrates around the solution of (\ref{basispursuit}). If the minimizer
is unique and we have access to the marginals of $\mu$, we can therefore
solve (\ref{basispursuit}). Belief propagation 
provides low-complexity heuristics for
approximating such marginals.

In order to introduce  belief propagation, 
consider the factor graph $G= (V,F,E)$ with variable nodes $V=[N]$,
factor nodes $F=[n]$ and edges $E=[N]\times [n] = \{(i,a):\,i\in [N],\,
a\in[n]\}$. Hence $G$ is the complete bipartite graph
with $N$ variable nodes and $n$ factor nodes.
It is clear that the joint distribution (\ref{eq:JointDistribution})
is structured according to this factor graph. 
Associated with the edges of this graph are the belief propagation
messages $\{\me_{i\to a}\}_{i\in V, a\in C}$ and $\{\mh_{a\to i}\}_{i\in V, a\in C}$.
In the present case, messages are probability measures over the
real line. The update rules for these densities are
\begin{eqnarray}
\me_{i \rightarrow a}^{t+1}(s_i) & \normeq &e^{-\beta |s_i|}
\prod_{b \neq a} \mh_{b \rightarrow i}^{t} (s_i)\, ,
\label{eq:SumProduct1}\\
\mh_{a \rightarrow i}^{t}(s_i)& \normeq & \int
\prod_{j \neq i} \nu_{j \rightarrow a}^{t}(s_i)\,\;
\delta_{\{y_a- (As)_a\}}\de s\, ,\label{eq:SumProduct2}
\end{eqnarray}
where a superscript denotes the iteration number and the symbol $\normeq$ denotes identity between probability
distributions up to a normalization constant\footnote{More precisely,
given two non-negative functions $p$, $q:\Omega \to\reals$
over the same space, we write $p(s)\normeq q(s)$
if there exists a positive constant
$a$ such that $p(s) = a\, q(s)$ for every $s\in\Omega$.}.
%
%
 \subsection{Large system limit}

A key remark is that in the large system limit, the messages
$\mh^{t}_{a\to i}(\, \cdot \,)$ are approximately
Gaussian densities with variances of order $N$, and the messages
$\me^{t}_{i\to a}(\, \cdot\, )$ are accurately approximated by
the product of a Gaussian and a Laplace density. We state this fact formally below. Recall that, given two measure 
$\mu_1$ and $\mu_2$ over $\reals$, their Kolmogorov distance is given by
$||\mu_1-\mu_2||_{\rm K} \equiv \sup_{a\in\reals}
|\mu_1(-\infty,a] - \mu_2(-\infty,a]|$.

The first Lemma is an estimate of the messages $\mh_{a\to i}^t$.
\begin{lemma} \label{lemma:SumProd1}
Let $x_{j \rightarrow a}^t$ and $(\tau_{j \rightarrow a}^t/\beta)$ be,
respectively, the mean and the
variance of the distribution $\me_{j \rightarrow a}^t$.
Assume further $\int |s_j|^3 \de\me_{j \rightarrow a}^t(s_j)\le C_t$
uniformly in $N,n$. Then there exists a constant  $C'_t$ such that
\begin{align}
&||\mh_{a\to i}^t-\ph_{a\to i}^t||_{\rm K}  \le \frac{C'_t}
{N^{1/2}(\htau^t_{a\to i})^{3}}\, ,\nonumber \\
&\ph_{a \rightarrow i}^{t}(\de s_i)  \equiv \sqrt{\frac{\beta A_{ai}^2}
{2\pi\htau_{a\to i}^t}}
\; \exp\left\{\frac{\beta}{2\htau_{a\to i}^t}
(A_{ai}s_i- z_{a \rightarrow i}^t)^2
\right\}\; \de s_i\; ,\label{eq:PhDef}
\end{align}
where the distribution parameters are given by
\begin{eqnarray}
z_{a \rightarrow i}^t \equiv y_a - \sum_{j \neq i} A_{aj} x_{j \rightarrow a}^t,
\;\;\;\;\;\;
\htau^t_{a\to i}
\equiv \sum_{j \neq i} A_{aj}^2 \tau_{j \rightarrow a}^t.\label{eq:htau}
\end{eqnarray}
\end{lemma}
Motivated by this lemma, we consider the computation of the means
and the variances of the messages  $\me_{i \rightarrow a}^{t+1}(s_i)$.
It is convenient to introduce a family of densities
\begin{eqnarray}
f_{\beta}(s;x,b) \equiv \frac{1}{z_{\beta}(x,b)}
\;\exp\Big\{-\beta|s|-\frac{\beta}{2b}(s-x)^2\Big\}\, .
\end{eqnarray}
Also let $F_{\beta}$ and $G_{\beta}$ denote its mean and variance, i.e.,
\begin{eqnarray}
\F_{\beta}(x;b)  \equiv \E_{f_{\beta}(\,\cdot\,;x,b)}(Z)\,,\;\;
\G_{\beta}(x;b)  \equiv \Var_{f_{\beta}(\,\cdot\,;x,b)}(Z)\, .
\end{eqnarray}
>From Eq.~(\ref{eq:htau}), we expect $\htau_{i\to a}^t$ to concentrate tightly. Therefore we assume that it is
independent of the edge $(i,a)$.
\begin{lemma}\label{lemma:SumProd2}
Suppose that at iteration $t$, the messages from the factor nodes to the
variable nodes are $\mh_{a \rightarrow i}^{t} = \ph_{a \rightarrow i}^{t}$,
with $\ph_{a \rightarrow i}^{t}$ defined as in Eq.~(\ref{eq:PhDef})
with parameters $z^{t}_{a\to i}$ and $\htau^{t}_{a\to i} = \htau^t$.
Then at the next
iteration we have
\begin{align}
\me_{i\to a}^{t+1}(s_i) &= \pe_{i\to a}^{t+1}(s_i)\,\{1+O(s_i^2/n)\}\, ,\nonumber \\
\pe_{i\to a}^{t+1}(s_i) &\equiv
f_{\beta}(s_i;\sum_{b \neq a} A_{bi}z_{b \rightarrow i}^t ,\htau^{t})\, . \nonumber
\label{eq:Me}
\end{align}
The mean and the variances of these messages are given by
\begin{align*}
x_{i \rightarrow a}^{t+1} &= \F_{\beta}(\sum_{b \neq a} A_{bi}z_{b \rightarrow i}^t ; \htau^t), \\
\tau_{i \rightarrow a}^t & = \beta
\, \G_{\beta}\Big(\sum_{b \neq a} A_{bi}z_{b \rightarrow i}^t ; \htau^t\Big).
\end{align*}
\end{lemma}
%
%
\subsection{Large $\beta$ limit}

In the limit $\beta \rightarrow \infty$, we can simplify the functions $\F_{\beta}$ and $\G_{\beta}$.
 Consider the soft thresholding function 
$\eta(x;b)= \textmd{sign}(x)(|x|-b)_+$.
It is well known that this admits the alternative characterization
\begin{eqnarray}
\eta(x;b) = \argmin_{s\in\reals}\left\{|s| + \frac{1}{2b}(s-x)^2\right\}\, .
\end{eqnarray}
In the $\beta\to\infty$ limit, the integral that defines $\F_{\beta}(x;b)$
is dominated by the maximum value of the exponent, that corresponds to
$s_* = \eta(x;b)$ and therefore  $\F_{\beta}(x;b)\to  \eta(x;b)$.
The variance (and hence the function $\G_{\beta}(x;b)$) can be estimated
by approximating the density $f_{\beta}(s;x,b)$ near $s_*$.
Two cases can occur. If $s_*\neq 0$, then a
Gaussian approximation holds and $\G_{\beta}(x;b)=\Theta(1/\beta)$.
On the other hand, if $s_*= 0$,  $f_{\beta}(s;x,b)$ can be approximated by a
Laplace distribution, leading to $\G_{\beta}(x;b)=\Theta(1/\beta^2)$
(which is negligible).
We summarize this discussion in the following.
\begin{lemma}\label{lemma:LargeBeta}
For bounded $x,b$, we have
\begin{eqnarray*}
\lim_{\beta\to\infty}\F_{\beta}(x;\beta) = \eta(x;b)\, ,\;\;\;\;\;
\lim_{\beta\to\infty}\beta\,\G_{\beta}(x;\beta) = b\,\eta'(x;b)\, .
\end{eqnarray*}
\end{lemma}

Lemmas \ref{lemma:SumProd1},\ref{lemma:SumProd2}, and
\ref{lemma:LargeBeta} suggest the 
following equivalent form for the message passing algorithm
(for large $\beta$):
\begin{align}
x_{i \rightarrow a}^{t+1} &= \eta
\Big(\sum_{b \neq a} A_{bi}z_{b \rightarrow i}^t ; \htau^t \Big)\, , \\
z_{a \rightarrow i}^t &\equiv y_a - \sum_{j \neq i} A_{aj} x_{j \rightarrow a}^t,
\label{eq:UpdateBetaInf1}\\
\htau^{t+1}
& =   \frac{\htau^t}{N\delta}
\, \sum_{i=1}^N\eta'\Big(\sum_{b} A_{bi}z_{b \rightarrow i}^t ; \htau^t\Big)\, .
\label{eq:UpdateBetaInf2}
\end{align}
%
%
%
\subsection{From message passing to AMP}

The updates in Eqs.~(\ref{eq:UpdateBetaInf1}),
(\ref{eq:UpdateBetaInf2}) are easy to implement but nevertheless the overall
algorithm is still rather complex because it requires to
track $2nN$ messages.
The goal of this section is to further simplify the
update equations. In order to justify the approximation we assume that the messages can be approximated as
$x_{i \rightarrow a}^t = x_i^t + \delta x_{i \rightarrow a}^t+ O(1/N)$,
$z_{a \rightarrow i}^t = z_a^t+ \delta z_{a \rightarrow i}^t+O(1/N)$,
with $\delta x_{i \rightarrow a}^t,\delta z_{a \rightarrow i}^t
= O(\frac{1}{\sqrt{N}})$ (here the $O(\,\cdot\,)$ errors are 
assumed uniform in the
choice of the edge).
We also consider a general message passing  algorithms of the
form
\begin{eqnarray}
x_{i \rightarrow a}^{t+1} = \eta_t
\Big(\sum_{b \neq a} A_{bi}z_{b \rightarrow i}^t
\Big)\, ,\;\;
z_{a \rightarrow i}^t \equiv y_a - \sum_{j \neq i} A_{aj} x_{j \rightarrow a}^t,
\label{eq:GeneralAlgo}
\end{eqnarray}
with $\{\eta_t(\,\cdot\,)\}_{t\in\naturals}$ a sequence of differendiable
nonlinear functions with bounded derivatives.
Notice that the algorithm derived at
the end of the previous section, cf. Eqs.~(\ref{eq:UpdateBetaInf1}),
Eqs.~(\ref{eq:UpdateBetaInf2}), is of this form,
albeit with $\eta_t$ non-differentiable at $2$ points.
But this does not change the result, as long as the
nonlinear functions are Lipschitz continuous. In the interest
of simplicity, we just discuss the differentiable model.
\begin{lemma}\label{lemma:AMP}
Suppose that the asymptotic behavior described in the paragraph above holds for the message
passing algorithm (\ref{eq:GeneralAlgo}). Then $x_i^{t}$ and $z_a^{t}$ satisfy
the following equations
\begin{align}
&x_i^{t+1} = \eta_t\Big( \sum_{a} A_{ia} z_a^t + x_i^t\Big) +o_N(1),  \nonumber \\
&z^t_a= \! y_a \! -\! \!  \sum_j A_{aj}x_j^t + \frac{1}{\delta} z_a^{t-1}\langle \eta_{t-1}'(  A^* z^{t-1}\! +  x^{t-1}) \rangle \! + \! o_N(1), \nonumber
\end{align}
where the $o_N(1)$ terms vanish as $N,n\to\infty$.
\end{lemma}
As a consequence, the resulting algorithm can be written in the vector notation as follows:
\begin{align}
x^{t+1}&= \eta( A^* z^t+x^t;\htau^t)\, , \label{eq:AMP1} \\
z^t&= y-A x^t+\frac{1}{\delta}z^{t-1}  \langle \eta'(A^* z^{t-1}+  x_i^{t-1};\htau^{t-1}) \>\, ,\label{eq:AMP2}
\end{align}
where $\langle \,\cdot\, \rangle$ denotes the average of 
a vector.

We also get the following recursion for $\hat{\tau}$:
\begin{eqnarray}
\htau^{t}= \frac{\htau^{t-1}}{\delta}
\<\eta'(A^* z^{t-1}+ x^t;\htau^{t-1}) \>\, .\label{eq:FinalTauRecursion}
\end{eqnarray}
%
%
\subsection{Comments}

\emph{Threshold level.} The derivation presented here yields a
`parameter free' algorithm. The threshold level $\htau^t$
is updated by the recursion in Eq. (\ref{eq:FinalTauRecursion}).
One could take the alternative point of view that $\htau^t$
is a parameter to be optimized. This point of
view was adopted in \cite{DMM,DMM2}. It is expected that the
two points of view coincide in the large system limit, but
it might be advantageous to consider a general sequence of
thresholds.

\emph{Mathematical derivation of AMP.}
We showed that in a specific limit (large systems, and large $\beta$)
the sum-product update rules can be significantly simplified
to (\ref{eq:AMP1}), (\ref{eq:AMP2}). We should emphasize
 that our results concern just a single step of the
iterative procedure. As such they do not prove that the sum-product 
messages are carefully
tracked by Eqs.~(\ref{eq:AMP1}), (\ref{eq:AMP2}). In principle it 
could be that the error terms in our approximation,
while negligible at each step, conjure up to become large
after a finite number of iterations. We do not expect this to be the
case, but it is nevertheless an open mathematical problem.
%
%
\section{AMP for  BPDN/Lasso}
\label{sec:AMPsoft}

Another popular reconstruction procedure in compressed sensing
is the following problem
\begin{eqnarray}
\mbox{minimize}\; \lambda\|s\|_1 + \frac{1}{2} \|y- As\|_2^2.\label{eq:Lasso2}
\end{eqnarray}
The derivation of the corressponding AMP is similar to the one
in the previous section. We therefore limit ourself to mentioning
a few differences.

As before we define a joint density distribution on the variables
$s=(s_1, \ldots, s_N)$
\begin{align*}
\mu(\de s)= \frac{1}{Z} \prod_{i=1}^N \exp(- \beta \lambda |s_i|)
\prod_{a=1}^{n}  \exp\Big\{-\frac{\beta}{2} (y_a- (As)_a)^2\Big\}\; \de s\, .
\end{align*}
The mode of this distribution
coincides with the solution of the problem (\ref{eq:Lasso2}) and
the distribution concentrates on its mode as $\beta\to\infty$.
The sum-product algorithm is
\begin{align*}
\me_{i \rightarrow a}^{t+1}(s_i) &\normeq \exp(-\beta \lambda |s_i|)
\prod_{b \neq a} \me_{b \rightarrow i}^t(s_i), \\
\mh_{a \rightarrow i}^{t}(s_i) & \normeq  \int
\exp\Big\{-\frac{\beta}{2}(y_a- (As)_a)^2\Big\} \, \prod_{j \neq i} \de
\me_{j \rightarrow a}^t(s_j)\,  .
\end{align*}

Proceeding as above, we derive an asymptotically 
equivalent form of the belief propagation for $N \rightarrow \infty$ and 
$\beta \rightarrow \infty$.
We get the following algorithm in the vector notation:
\begin{align}
x^{t} & =  \eta(x^t+ A^*z^t; \lambda+ \gamma^t )\, , \\
z^{t+1} & =  y - Ax^t + \frac{1}{\delta} z^t \langle \eta'(x^{t-1}+ A^* z^{t-1}),  \rangle
\end{align}
which generalize Eqs.~(\ref{eq:AMP1}) and (\ref{eq:AMP2}).
The threshold level is computed iteratively as follows
\begin{eqnarray}
\gamma^{t+1} = \frac{\lambda + \gamma^t}{\delta}
\<\eta'(A z^t + x^t; \gamma^t+ \lambda)\>\, .
\end{eqnarray}
Notice that the only deviation from the algorithm in the previous section
is in the recursion for the threshold level.
%
%
\section{AMP for reconstruction with prior information}
\label{sec:AMPBayes}

In the two cases we discussed so far, the distribution of the signal 
$s_o$ was not known. This is a very natural and practical assumption. 
Nevertheless, it might be possible in specific scenarios
to estimate the input distribution. This extra information may be used to 
improve the recovery algorithms. Also, the case of known
signal distribution provides a benchmark for the other approaches. 
In this section we define a very simple iterative theresholding algorithm
for these situations.

Let $\alpha = \alpha_1\times \alpha_2 \dots \times \alpha_N$ be the joint probability distribution of the variables $s_1,s_2, \ldots, s_N$.
It is then natural to consider the distribution
\begin{eqnarray*}
\mu(\de s)= \frac{1}{Z}  \prod_{a=1}^n \exp\Big\{-\frac{\beta}{2}(y_a-
(As)_a)^2\Big\}
\prod_{i=1}^N \alpha_i(\de s_i)\, ,
\end{eqnarray*}
since $\mu$ is the \emph{a posteriori} distribution of $s$, when $y = As+w$ is observed. Here, $w$ is a noise vector with i.i.d. normal entries and is independent of $s$. The sum-product update rules are
\begin{align*}
\me_{i \to a}^{t+1}(\de s_i) &\normeq  \prod_{b \neq a} \mh_{b \rightarrow i}^t(s_i) \,  \alpha_i(\de s_i)\, ,  \\
\me_{a \to i}^{t}(s_i) &\normeq \int
\exp\Big\{-\frac{\beta}{2}(y_a- (As)_a)^2\Big\}
 \prod_{j \neq i} \nu_{j \rightarrow a}^t(\de s_j) \, . 
\end{align*}
Notice that the above update rules are well defined. At each
iteration $t$, the message
$\me_{i \to a}^{t+1}(\de s_i)$ is a probability measure
on $\reals$, and the first equation gives
its density with respect to $\alpha_i$.
The message $\me_{a \to i}^{t}(s_i)$ is instead a non-negative
measurable function (equivalently, a density) given by
the second equation. Clearly the case studied in the previous section
corresponds to $\alpha_i \normeq \exp(-\beta |s_i|)$.

In order to derive the simplified version of the message passing algorithm,
we introduce the following family of measures over $\reals$
\begin{eqnarray}
f_{i}(\de s;x,b) \equiv \frac{1}{z_{\beta}(x,b)}
\;\exp\Big\{-\frac{\beta}{2b}(s-x)^2\Big\}\, \alpha_i(\de s)\, ,
\end{eqnarray}
indexed by $i\in [N]$, $x\in\reals$, $b\in\reals_+$
($\beta$ is fixed here). The mean and the variance of this distribution
define the functions
(here $Z \sim f_{i}(\,\cdot\,;x,b)$)
\begin{eqnarray}
\F_{i}(x;b)  \equiv \E_{f_{i}(\,\cdot\,;x,b)}(Z)\,,\;\;\;\;
\G_{i}(x;b)  \equiv \Var_{f_{i}(\,\cdot\,;x,b)}(Z)\, .
\end{eqnarray}
These functions have a natural estimation-theoretic interpretation.
Let $X_i$ be a random variable with distribution $\alpha_i$,
and assume that $\tY_i = X_i+W_i$ is observed with $W_i$ gaussian noise
with variance $b/\beta$. The above functions are --respectively--
the conditional expectation and conditional variance of $X_i$,
given that $\tY_i=x$:
\begin{eqnarray*}
\F_{i}(x;b)  = \E(X_i| \tY_i= x)\,,\;\;\;\;
\G_{i}(x;b)  = \Var(X_i|\tY=x)\, .
\end{eqnarray*}

The approach described in Section \ref{sec:AMPhard}
yields the following AMP (in vector notation)
\begin{eqnarray}
x^{t} & = & \F(x^t+ A^*z^t; \lambda+ \gamma^t ), \\
z^{t+1} & = & y - Ax^t + \frac{1}{\delta} z^t \langle \F'(x^{t-1}+ A^* z^{t-1})  \rangle\, .
\end{eqnarray}
Here, if $x\in\reals^N$, $\F(x;b)\in\reals^N$ is the vector
$\F(x;b) = (\F_1(x_i;b),\F_2(x_2;b),\dots,\F_N(x_N;b))$.
Analogously $\F'(x) = (\F_1'(x_i;b),\F_2'(x_2;b),\dots,\F_N'(x_N;b))$
(derivative being taken with respect to the first argument).
Finally, the threshold level is computed iteratively as follows
\begin{eqnarray}
\gamma^{t+1} = \frac{1}{\delta}
\<\G(A z^t + x^t; \gamma^t+ \lambda)\>\, .
\end{eqnarray}
%
%
%
\subsection{Comments}

The AMP algorithm described in this section is marginally more
complex than the ones in the previous sections.
The main difference is that the soft thresholding
function $\eta(\,\cdot\,)$  is replaced with the conditional expectation
$\F(\,\cdot\,)$. While the latter does not admit, in general,
a closed form expression, it is not hard to construct
accurate approximations that are easy to evaluate.
%
%
\section{Related work}
\label{sec:Related}

In this section we would like to clarify the relation 
of the present approach with earlier results in the literature.
Each of these lines of work evolved from different 
motivations, and there was so far little -- if any -- contact between them.

\subsection{Other message passing algorithms}

The use of message passing algorithms for compressed sensing problems was
suggested before, see for instance \cite{BaronBP}. However such a proposal 
faced two major difficulties.

\vspace{0.1cm}

\noindent\emph{(1)} According to the standard prescription,
messages used in the the sum-product algorithm should be 
probability measures over the real line $\reals$, cf.  
Eqs.~(\ref{eq:SumProduct1}), (\ref{eq:SumProduct2}). This 
is impractical from a computational point of view.
(A low complexity message-passing algorithm 
for a related problem was used in \cite{Braid}).

\vspace{0.1cm}

\noindent\emph{(2)} The factor graph on which the sum-product algorithm
is run is the complete bipartite graph with $N$ variable nodes,
and $n$ function nodes. In other words, unless the underlying matrix is
sparse \cite{Indyk}, the graphical model is very dense.
This requires to update $Nn$ messages per
iteration, and each message update depend on $N$ or $n$ input messages. 
Again this is very expensive computationally.
 
\vspace{0.1cm}

The previous pages show that problem $(2)$ does not 
add to $(1)$, but in fact solves it!
Indeed, the high density of the  graph leads to approximately Gaussian
messages from factor nodes to variable nodes, via central limit theorem. 
Gaussian messages are in turn parametrized by two numbers: mean and 
variance. It turns out that is is sufficient to keep track only of the 
means, again because of the high density.

Problem $(2)$ is also solved by the high density nature
of the graph, since all the messages departing from the same node
of the graph are very similar with each other. 

One last key difficulty with the use of belief propagation
in compressed sensing was

\vspace{0.1cm}

\noindent\emph{(3)} The use of belief propagation 
requires to define a prior on the vector $s_o$. For most applications, 
no good prior is available. 

\vspace{0.1cm}

The solution of this problem lies in using a Laplace prior 
as in Eq.~(\ref{eq:JointDistribution}). A first justification of this
choice lies in the fact that, as $\beta\to\infty$, the resulting probability
measure concentrates around its mode, that is the solution of
the basis pursuit problem (\ref{basispursuit}). A deeper
reason for this choice is that it is intimately related to the 
soft threshold non-linearity $\eta(x;\theta)$, which is 
step-by-step optimal in a minimax sense \cite{DMM,DMM2}.

%
%
\subsection{Historical background and statistical physics}

There is a well studied connection between statistical physics
techniques and message passing algorithms \cite{MezardMontanari}. 
In particular, the sum-product algorithm corresponds to 
the Bethe-Peierls approximation  in statistical physics, and its
fixed points are stationary points of the Bethe free energy.
In the context of  spin glass theory, the
Bethe-Peierls approximation is also referred to as the 
`replica symmetric cavity method'. 

The Bethe-Peierls approximation postulates a set 
of non-linear equations on quantities that are equivalent to the
sum-product messages, and which are in correspondence with local
marginals.  
In the special cases of 
spin glasses on the complete graph (the celebrated Sherrington-Kirkpatrick 
model), these equations reduce to the so-called TAP equations,
named after Thouless, Anderson and Palmer who first used them
\cite{TAP}.

The original TAP equations where a set of non-linear 
equations for local magnetizations (i.e. expectations of a single variable).
Thouless, Anderson and Palmer first recognized that naive mean field 
is not accurate enough in the spin glass model, and corrected it 
by adding the so called Onsager reaction term that is analogous 
to the last term in Eq.~(\ref{eq:AMP2}).
More than 30 years after the original paper, a complete mathematical 
justification of the TAP equations remains an 
open problem in spin glass theory,
although important partial results exist \cite{Talagrand}.
While the connection between belief propagation and 
Bethe-Peierls approximation stimulated a considerable 
amount of research \cite{Yedidia}, the algorithmic uses of TAP equations
have received only sparse attention. Remarkable exceptions
include \cite{OpperWinther,Kabashima,NeirottiSaad}.
%
%
\subsection{State evolution and replica calculations}

In the context of coding theory, message passing algorithms 
are analyzed through density evolution \cite{RiU}. 
The common justification for density evolution is that the underlying
graph is random and sparce, and hence converges locally to a tree
in the large system limit. In the case of trees density
evolution is exact, hence it is asymptotically exact for sparse 
random graphs.

State evolution is the analog of density evolution in the 
case of dense graphs.
For definitions and results on state evolution we refer to the
\cite{DMM,DMM2}.
The success of state evolution cannot be ascribed to the locally
tree-like structure of the graph, and calls for new mathematical ideas.

The fixed points of state evolution describe the 
output of the corresponding AMP, when the latter 
is run for a sufficiently large number of iterations
(independent of the dimensions $n,N$). 
It is well known, within statistical mechanics \cite{MezardMontanari},
that the fixed point equations do indeed coincide with 
the equations obtained through a completely different 
non-rigorous approach, the \emph{replica method} 
(in its replica-symmetric form). 
This is indeed an instance
of a more general equivalence between replica and cavity methods.

During the last few months, several papers investigated
compressed sensing problems using the replica method
\cite{Goyal,KabashimaTanaka,BaronGuoShamai}.
In view of the discussion above, it is not surprising
that these results can be recovered from the state evolution formalism
put forward in \cite{DMM}. Let us mention that the latter 
has several advantages over the replica method:
$(1)$ It is more concrete, and its assumptions can be 
checked quantitatively through simulations; $(2)$
It is intimately related to efficient message passing algorithms;
$(3)$ It actually allows to predict the performances of these 
algorithms.

\section*{Acknowledgment}

This work was partially supported by
a Terman fellowship, the NSF CAREER award CCF-0743978 and
the NSF grant DMS-0806211.

\end{document}